# $d^0$ Ferromagnetism in Ag-doped Monoclinic ZrO$_2$ Compounds

L. Chouhan[1], G. Bouzerar[2] and S. K. Srivastava[1*]

[1]Department of Physics, Central Institute of Technology Kokrajhar, Kokrajhar-783370, India
[2]CNRS et Université Claude Bernard Lyon 1, F-69622, Lyon, France

[*]**Corresponding Author E-mail:** sk.srivastava@cit.ac.in

**Abstract**

Recently $d^0$ or intrinsic ferromagnetism was believed to provide an alternative pathway to transition metal induced ferromagnetism in oxide. In pursuit of augmenting the area of $d^0$ ferromagnetism; we have undertaken to study the crystal structure and magnetic properties of Ag-doped ZrO$_2$ compounds. Polycrystalline samples of Zr$_{1-x}$Ag$_x$O$_2$ (with x=0, 0.02, 0.04, 0.06 and 0.08) were prepared by solid-state reaction route. All the prepared compounds are found to crystallize in monoclinic symmetry of ZrO$_2$. In our study, pure ZrO$_2$ compound exhibits paramagnetic behavior. However, the Ag-doped ZrO$_2$ compounds exhibit ferromagnetic to paramagnetic transition. The Curie temperature ($\theta_C$) was found to increase from 28.7 K for x=0.02 to 173.2 K for x= 0.08 doped ZrO$_2$. Thus, the introduction of Ag in ZrO$_2$ induces ferromagnetism with a large $\theta_C$. The measurements of hysteresis curves indicate that Ag doped ZrO$_2$ compounds exhibit hysteresis loops with a coercivity of around 1350 Oe. Moreover, increase in Ag concentration resulted increase in the value of saturation magnetization (M$_S$); the maximum value of M$_S$ was recorded as 0.01 $\mu_B$/Ag ion for x= 0.06 sample. The sintering of sample at high temperature (1350$^0$C) diminishes the ferromagnetism and it leads to paramagnetic behaviour.

# 1. Introduction:

Over last couple of decades, there has been persistent effort by researchers to explore new materials which can be integrated for spintronics devices, such as giant magneto-resistance sensors, magneto-resistive random-access memories and storage media [1]. Spintronics devices, where both charge as well as spin are utilized, have been projected to have plenty of advantages over prevalent semiconductor devices, such as faster data processing, non-volatility, low power consumption and increased storage density [1, 2]. Among the various explored materials for spintronics devices, transition metal (TM) doped semiconducting oxide materials were studied extensively. Many TM-doped oxides, such as ZnO, $SnO_2$ and $TiO_2$ have been reported to exhibit room temperature ferromagnetism (RTFM) [3-11]. The magnetic coupling through exchange interaction of delocalized 3d electrons via structural defects (oxygen vacancies, $V_o$) has been put forward as possible explanation of the ferromagnetism observed in TM-doped oxides [10]. Although, the experimental search for ferromagnetism (FM) in TM-doped oxide materials has been continued to be dogged, but it has not yet resulted in reproducible and homogeneous magnetic materials and there is a debate whether RTFM is intrinsic or due to the magnetic ion cluster [11-13].

In order to get a clean material exhibiting RTFM, several other types of materials were studied and explored. In addition to TM-doped oxide materials, unexpected ferromagnetism has been reported or predicted in several pure oxides like $HfO_2$, CaO, ZnO, $ZrO_2$, $TiO_2$, MgO, $SnO_2$ and, even in $CaB_6$ [14-20]. Such unexpected ferromagnetism was termed as $d^0$ ferromagnetism or intrinsic ferromagnets i.e. the materials that do not contain magnetic impurities. Thus, the $d^0$ or intrinsic ferromagnetism was believed to provide an alternative pathway to TM-induced ferromagnetism. The origin of the magnetism in these materials was considered due to point defects such as cation vacancies, which induces a local magnetic moment on the neighboring oxygen atoms [15, 20]. To circumvent the difficulties of defect control, an alternative way, which consists of the substitution of non-magnetic elements in dioxides such as $AO_2$ (*A*=Ti, Zr, or Hf), was proposed [21]. Following to this idea, many *ab-initio* studies have predicted ferromagnetism with high Curie Temperature ($\theta_C$) in several non-magnetic elements doped oxides, such as K-$SnO_2$ [22], Ag-$SnO_2$ [23], Mg–$SnO_2$ [24], anatase Li-$TiO_2$ [25], rutile K–$TiO_2$ [26], V-$TiO_2$ [27], K–$ZrO_2$ [16, 26]. Experimentally, $d^0$ magnetism were observed in several non-magnetic elements doped oxides such as; alkali metal doped ZnO [28-30]; Cu doped $TiO_2$ prepared in thin film form [31, 32], C-doped $TiO_2$ prepared by solid state route [33], K-SnO$_2$ [34], Li-SnO$_2$ [35], K-TiO$_2$ [36], Cu-ZnO [37] and Na-SnO$_2$ [38].

Recently Zirconium dioxide (ZrO$_2$) was projected as one of the promising candidates exhibiting $d^0$ ferromagnetism. ZrO$_2$ is a multipurpose material used in various scientific & technological applications due to its high dielectric constant, ionic conductivity, wide optical band gap, high chemical and thermal stabilities, low optical loss and high transparency [39-41]. It can crystallize in three different forms i.e. monoclinic (space group: P2$_1$/c), tetragonal (space group: P4$_2$/nmc) and cubic (space group: Fm3m) structures [39-40]. The monoclinic phase of zirconia is usually thermodynamically stable up to 1400 K. The tetragonal and cubic phases of ZrO$_2$ can be stabilized either by heating at very high temperature (1480 <$T$ <2650 K for tetragonal phase and, $T$ >2650 K for cubic phase) or by the addition of another cation such as Ca$^{2+}$ or Y$^{3+}$ [40]. In fact, several recent reports indicate that pure ZrO$_2$ exhibits room temperature d$^0$ ferromagnetism and it is related to the presence of oxygen vacancies or structural defects. Some of these studies also show that the crystallographic phase is very important in this context, with reports of ferromagnetism more common for tetragonal ZrO$_2$ structures [42-44]. The theoretical studies predicted high-temperature ferromagnetism in TM-doped cubic zirconia [45] as well as non-magnetic element doped zirconia such as K [16, 26] and V [46] in ZrO$_2$. However, it was predicted that doping with Cu [47], Cr [45] or Ca [16] in ZrO$_2$ can result in paramagnetism, antiferromagnetic or non-magnetic ground states, respectively.

Although, Ag-doped ZrO$_2$ was attempted by researchers for different kind of application such as in resistive switching, soot oxidation and opto-electronics [48-50] but, there exists no report on study of magnetic properties of Ag-doped ZrO$_2$. In pursuit of augmenting the research area of $d^0$ ferromagnetism, we have endeavoured a study on the crystal structure and magnetic properties of Ag-doped ZrO$_2$ compounds. We have chosen to prepare materials in bulk form at equilibrium conditions to diminish the uncertainties in fabrications and any inaccuracies in characterization.

## 2. Experimental Details

The polycrystalline samples of Zr$_{1-x}$Ag$_x$O$_2$ (x=0, 0.02, 0.04, 0.06 and 0.08) were prepared by solid-state reaction route. We used high-purity ZrO$_2$ and AgNO$_3$ as the starting materials for synthesis of our samples. The maximum amount of any kind of trace magnetic impurities in the starting materials was found to be less than 0.9 % ppm as mentioned by the supplier

ICP chemical analyses report. Pre-sintering of the prepared samples was performed in powder form at various temperatures, i.e. $200^0$C and $300^0$C for about 20 hours at each temperature. The samples were further annealed in pallet form at 500˚C for 30 hrs. One sample of $Zr_{0.94}Ag_{0.06}O_2$ was prepared at $1350^0$C to study the influence of sintering temperature. The crystal structure of the prepared samples was checked using X-ray diffractometer. The temperature (T) variation of magnetization (M) and, magnetization versus magnetic field (H) measurements were carried out using commercial SQUID magnetometer and vibrating sample magnetometer (VSM). The magnetic measurements were done with utmost care and repeated two times with different pieces of samples to guarantee the reproducibility of results

## 3. Results and Discussion

The crystal structure and phase purity of all Ag-doped $ZrO_2$ compounds were checked by X-ray diffractometer and the XRD patterns of these compounds are shown in Figure 1. All the XRD peaks could be indexed to monoclinic symmetry of $ZrO_2$. Within the instrumental limit of the X-ray diffractometer, the XRD patterns also indicate that samples are formed in single phase and no secondary phase is present. To gain insight into various crystal structural parameters, the refinement of the XRD patterns was performed with the help of the Fullprof program by employing the Rietveld refinement technique [51]. Figure 2 shows the typical Rietveld refinement of XRD patterns for $ZrO_2$ and $Zr_{0.94}Ag_{0.06}O_2$ compounds. It is seen that the experimental XRD data matches perfectly with the Rietveld software calculated XRD data. Figure 3 presents the variation of lattice parameters 'a' 'b', 'c' and cell volume (V) for $Zr_{1-x}Ag_xO_2$ (x=0, 0.02, 0.04, 0.06 and 0.08) compounds. For pure $ZrO_2$ compound, the lattice parameters are estimated to be a=5.1468Å, b= 5.2041Å, c=5.3198 Å and they are found to be comparable with the values reported in other work [21]. The lattice parameters and unit cell volume of the all Ag-doped $ZrO_2$ compounds are found to increase with the increase of Ag-doping. Doping of bigger $Ag^{1+}$ ion (ionic radii of 1.15 Å) into the $Zr^{4+}$ (ionic radii of 0.72 Å) is the likely cause of expansion of the lattice parameters and cell volume.

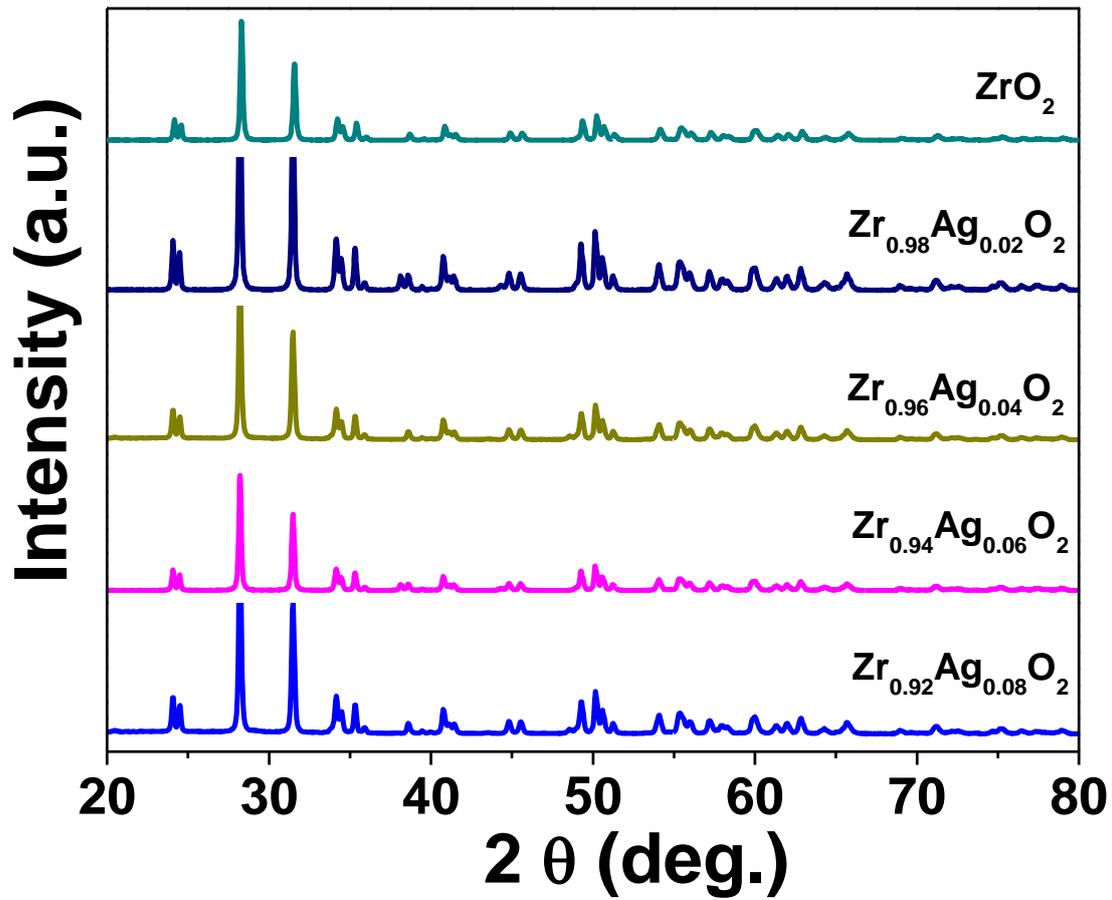

**Figure 1:** XRD patterns of $Zr_{1-x}Ag_xO_2$ (x=0, 0.02, 0.04, 0.06 & 0.08) compounds.

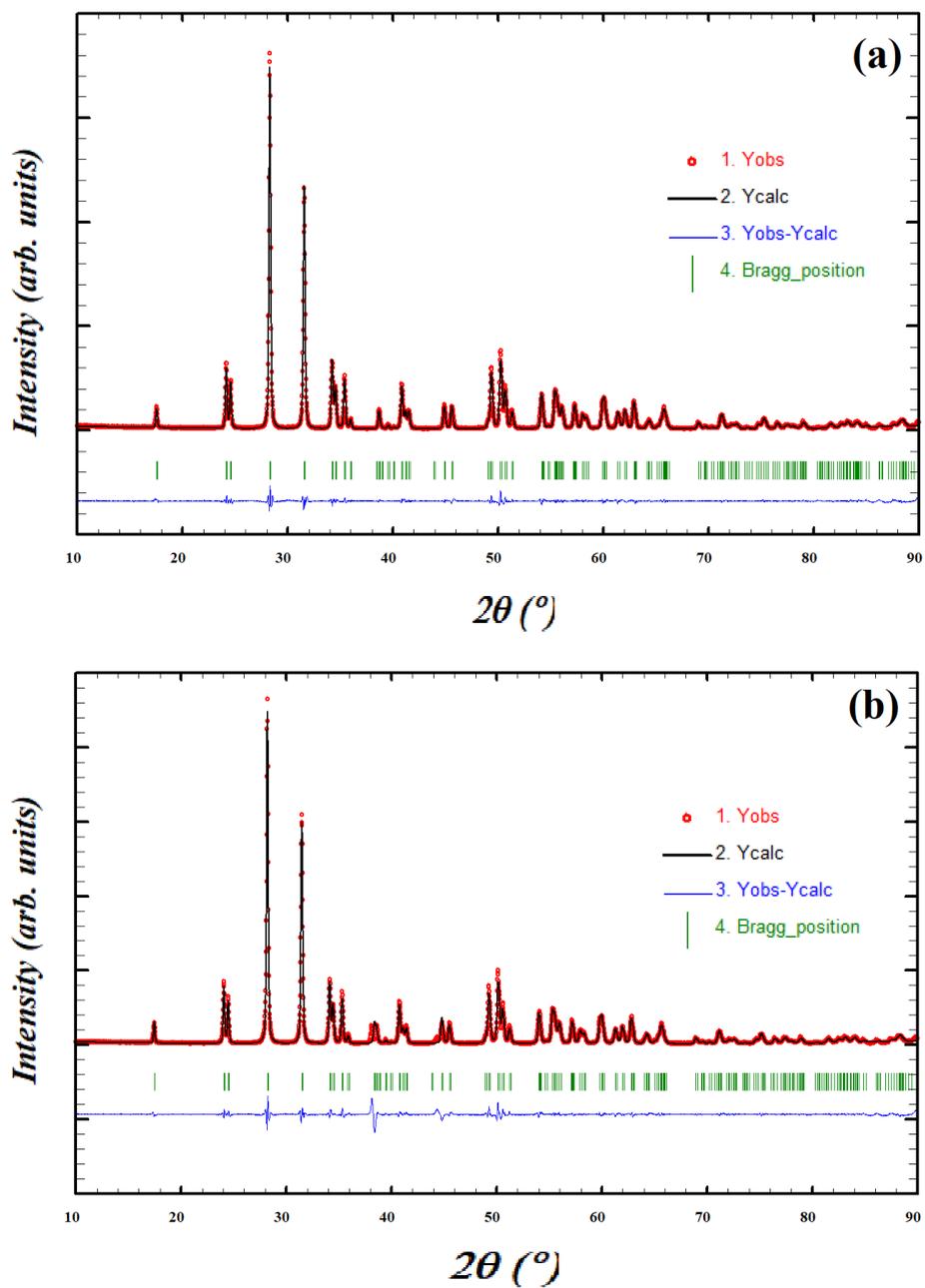

**Figure 2:** Refinement of XRD patterns for (a) $ZrO_2$ (b) $Zr_{0.94}Ag_{0.06}O_2$ compounds, obtained with the help of the Fullprof program by employing the Rietveld refinement technique.

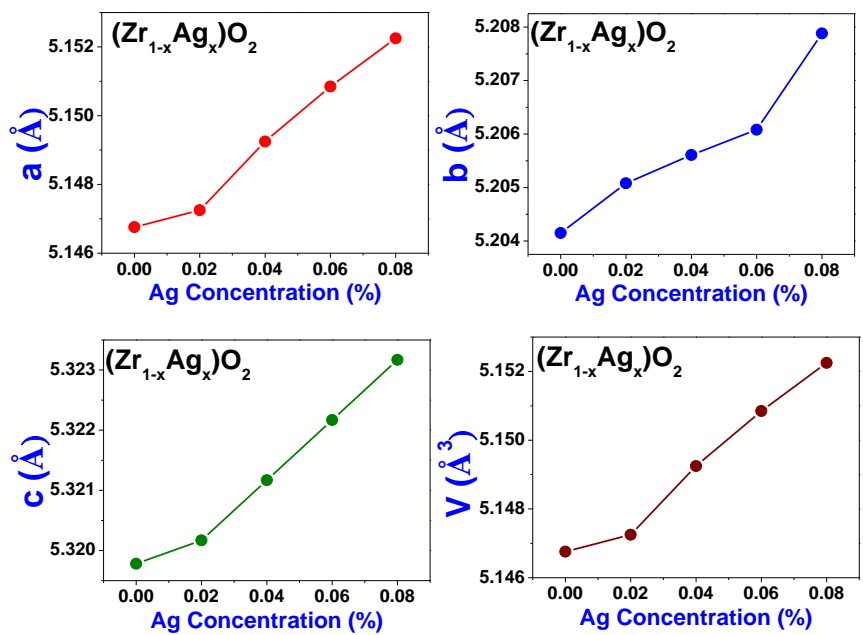

**Figure 3:** Variation of lattice parameters 'a' 'b', 'c' and cell volume (V) for $Zr_{1-x}Ag_xO_2$ (x=0, 0.02, 0.04, 0.06 and 0.08) compounds.

In order to explore the magnetic properties of these prepared compounds, the zero-field cooled (ZFC) magnetization curves as a function of temperature for all Ag-doped $ZrO_2$ compounds were measured under an applied field of 500 Oe using SQUID magnetometer. The temperature variation of magnetization as a function of temperature i.e. M-T curve for pure $ZrO_2$ compound shows a paramagnetic behavior of the sample, as depicted in Figure 4(a). In addition, the magnetization versus applied field measurement performed at 3K for pure $ZrO_2$ again indicates the paramagnetic nature of the compound as shown in figure 4(b). Thus, pure $ZrO_2$ compound is found to exhibit paramagnetic behavior. Figure 5 presents the temperature variation of magnetization curves for all Ag-doped $ZrO_2$ compounds. It is observed that the magnetization decreases with the increase of temperature throughout the measured temperature range for $Zr_{0.98}Ag_{0.02}O_2$ compound. However, higher Ag-doped compounds i.e. $Zr_{0.96}Ag_{0.04}O_2$ and $Zr_{0.94}Ag_{0.06}O_2$ compounds are found to exhibit a clear ferromagnetic to paramagnetic transition. Moreover, $Zr_{0.92}Ag_{0.08}O_2$ compound exhibits interesting feature. The measurement of ZFC M-T curve for this sample show that there is a low temperature antiferromagnetic (AFM) transition at ~55 K, followed by weak ferromagnetic to paramagnetic transition at ~165 K (as shown in the inset of Figure 5d). It indicates that weak ferromagnetic phase is superimposed with the dominating antiferromagnetic phase. To get further insight into the magnetic property of $Zr_{0.92}Ag_{0.08}O_2$ compound, we have measured the M-T curve under field cooled (FC) condition along with ZFC condition and the data are presented in Figure 5 (d). The measurements of ZFC and FC magnetization data for this sample indicate that ZFC and FC M-T curves coincide at low temperature and the magnitude of magnetization at AFM transition temperature of under FC condition has increased, indicating an antiferromagnetic interaction in the matrix. Similar observations have been reported in Fe-doped $SnO_2$ compounds [52]. To determine the ferromagnetic (FM) transition temperature ($T_C$), peaks observed in |dM/dT| versus temperature plot have been used. Here, $T_C$ value was taken as the minimum of |dM/dT| plot and fitting the curve with a Gaussian function. Typical plots of |dM/dT| versus temperature for x = 0.04 and 0.08 samples are shown in Figure 6. The $T_C$ values are obtained as 108.2, 192.4 and 172.6 K for x = 0.04, 0.06 and 0.08 samples respectively. Thus, the introduction of Ag in $ZrO_2$ gives rise to increase in ferromagnetic $T_C$. Moreover, the magnitude of magnetization is found to increase with Ag concentration in addition to increase in the FM $T_C$. The decrease of magnetization for x=0.08 sample is due to presence of competing AFM interaction. For the estimation of Curie-temperature ($\theta_C$), the paramagnetic region of ZFC M-T curve was analyzed using Curie-Weiss law, $\chi = C_0 x/(T - \theta_C)$. Typical plots of $1/\chi_{dc}$ versus temperature for x= 0.02, 0.04 and 0.08 samples are shown in Figure 7 along with Curie-Weiss law fitting and the estimated Curie-temperature ($\theta_C$) values are listed in Table 1. The value of Curie-temperature ($\theta_C$) is found to be 28.7, 126.8 and 163.2 K for x=0.02, 0.04 and 0.08 samples respectively. The positive values of $\theta_C$ indicate the FM interaction. The difference between $T_C$ and $\theta_C$ are mainly due to the observed broad magnetic transition. We could not fit data for x=0.06 sample due to non-availability of sufficient data in the paramagnetic region. For calculating effective paramagnetic moment ($\mu_{eff}$), the relation, $\mu_{eff} = \sqrt{3k_B C_0 x/N\mu_0 \mu_B^2}$ was used and the values were found to be 1.40 $\mu_B$/Ag ion, 2.27 $\mu_B$/Ag ion and 0.16 $\mu_B$/Ag ion for 2, 4 and 8 % Ag-doped samples respectively.

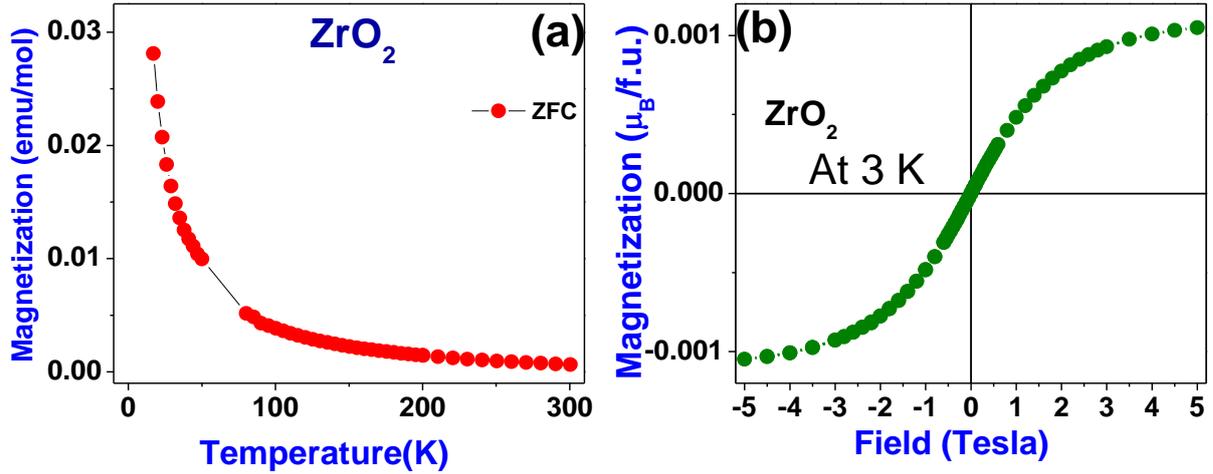

**Figure 4:** (a) M-T curve for pure $ZrO_2$ under a field of 0.05 T and (b) M-H loop obtained at 3 K for the $ZrO_2$ sample.

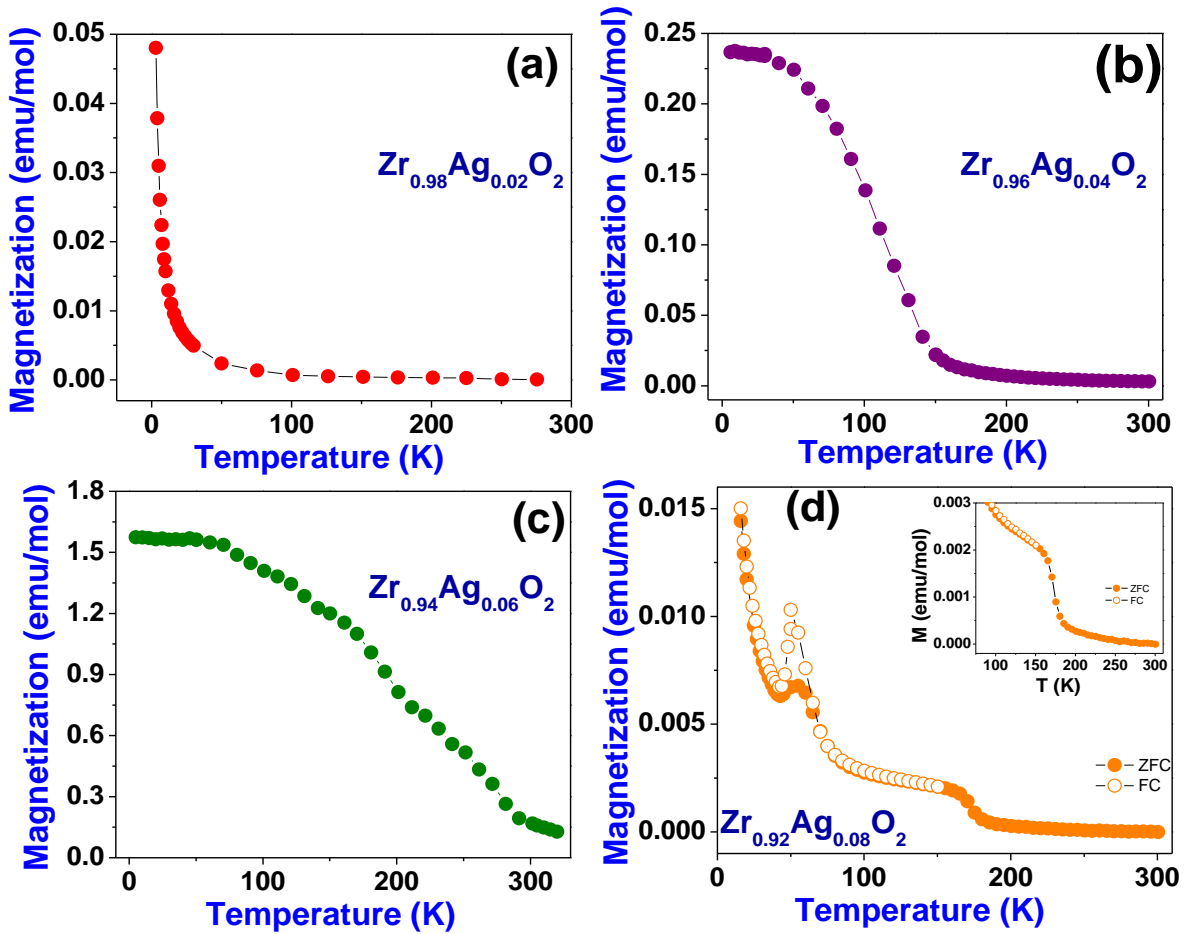

**Figure 5:** (a) Temperature variation of magnetization of Ag-doped $ZrO_2$ samples measured under an applied field of 0.05 T field.

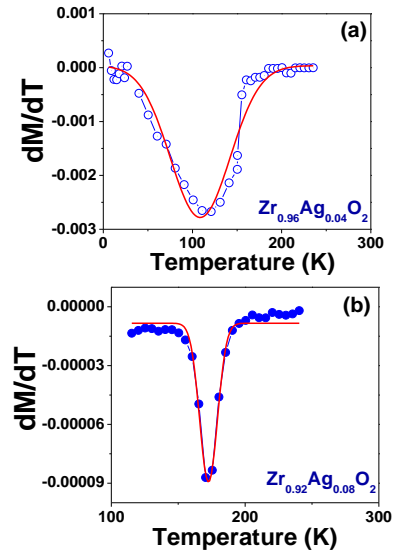

**Figure 6:** Temperature variation of |dM/dT| for $Zr_{0.96}Ag_{0.04}O_2$ and $Zr_{0.92}Ag_{0.08}O_2$.

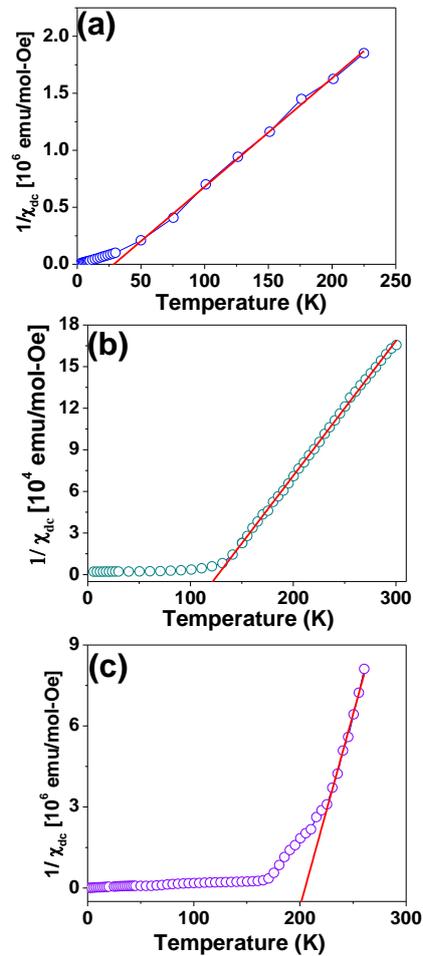

**Figure 7:** Temperature variation of inverse of susceptibility ($1/\chi_{dc}$) for $Zr_{1-x}Ag_xO_2$ with (a) x=0.02 (b) x=0.04 and (c) x=0.08 compounds. Solid lines represent fit to the Curie-Weiss law.

The field variation of magnetization measurements i.e. *M*-H curves at 3 K were performed for all Ag-doped $ZrO_2$ compounds and they are shown in Figure 8. From the curves, it is observed that the magnetization of these samples gets saturated at relatively larger applied field. One can see that the samples show a clear ferromagnetic behaviour at 3 K. The value of saturation magnetization ($M_S$) is found to be 0.003, 0.006, 0.009 and 0.004 $\mu_B$/Ag ion for x= 0.02, 0.04, 0.06 and, 0.08 respectively. The decrease in $M_s$ value for x=0.08 sample is possibly due to the presence of AFM interaction in the sample. Although, $Zr_{0.98}Ag_{0.02}O_2$ does not clearly exhibit any hysteresis loop but, higher Ag doped $ZrO_2$ compounds exhibit hysteresis loops with a coercivity of around 1350 Oe.

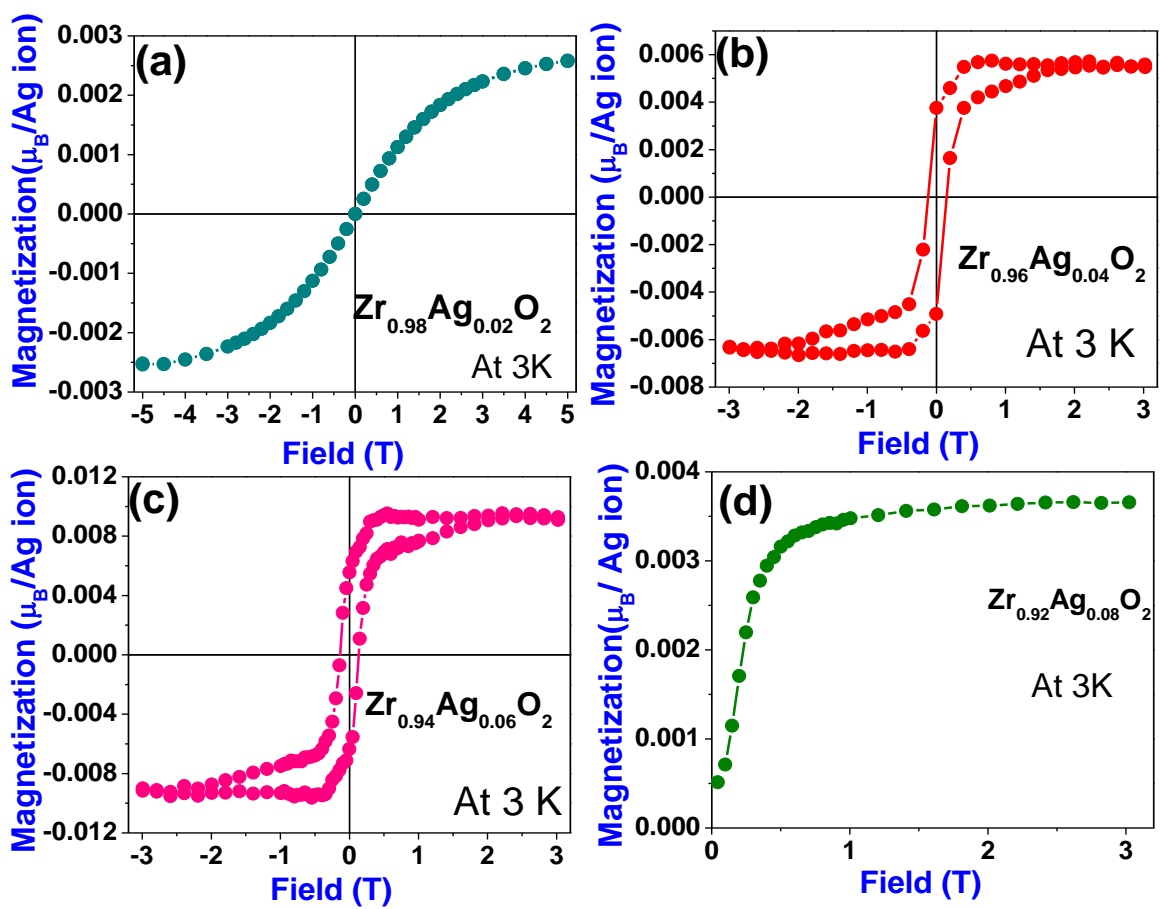

**Figure 8:** Field variation of magnetization for $Zr_{1-x}Ag_xO_2$ (x=0.02, 0.04, 0.06 and 0.08) compounds measured at 3 K.

**Table 1:** Parameters obtained from ZFC magnetization measurements of $Zr_{1-x}Ag_xO_2$ (x=0, 0.02, 0.04, 0.06, 0.08) compounds. Here, $T_C$ is ferromagnetic transition temperature. $\theta_C$ and $\mu_{eff}$ ($\mu_{B/ion}$) are Curie temperature and experimental effective magnetic moment respectively, obtained from the Curie-Weiss law fit of susceptibility data.

| Sample/ Parameters | x=0.02 | x=0.04 | x=0.06 | x=0.08 |
|---|---|---|---|---|
| $T_C$ (K) | -- | 108.6 | 192.4 | 172.6 |
| $\theta_c$ (K) | 28.7 | 126.8 | -- | 163.2 |
| $\mu_{eff}$ ($\mu_B$/Ag ion) | 1.40 | 2.27 | -- | 0.16 |
| $M_S$ ($\mu_B$/Ag ion) | 0.003 | 0.006 | 0.009 | 0.004 |

To study the influence of sintering temperature on the magnetic property, one sample of $Zr_{0.94}Ag_{0.06}O_2$ was prepared by sintering it at 1350ºC. The crystal structure from XRD patterns indicates that sample has been crystallized in monoclinic symmetry of $ZrO_2$ (not shown). Figure 9 shows the M-T and M–H (at 3K) curves measured for this sample. The measurement of M-T curve, as shown in Figure 9(a) indicates that it exhibits paramagnetic behaviour and it is further corroborated by M-H curve, as shown in Figure 9 (b). Moreover, from the M-H curve, it is observed that the value of magnetization decreases with increase in the sintering temperature. The saturation magnetization is two times larger for the sample prepared at 500 ºC, as compared with the sample prepared at 1350 ºC. These results suggest that oxygen vacancies present in the sample prepared at lower temperature lead to ferromagnetism, which get diminished in the sample prepared at high-temperature. High-temperature preparation results in the destruction of the ferromagnetic ordering and reduced magnetic moment, possibly due to decrease in oxygen vacancies. Similar observation was made for Cu doped $TiO_2$ [31-32].

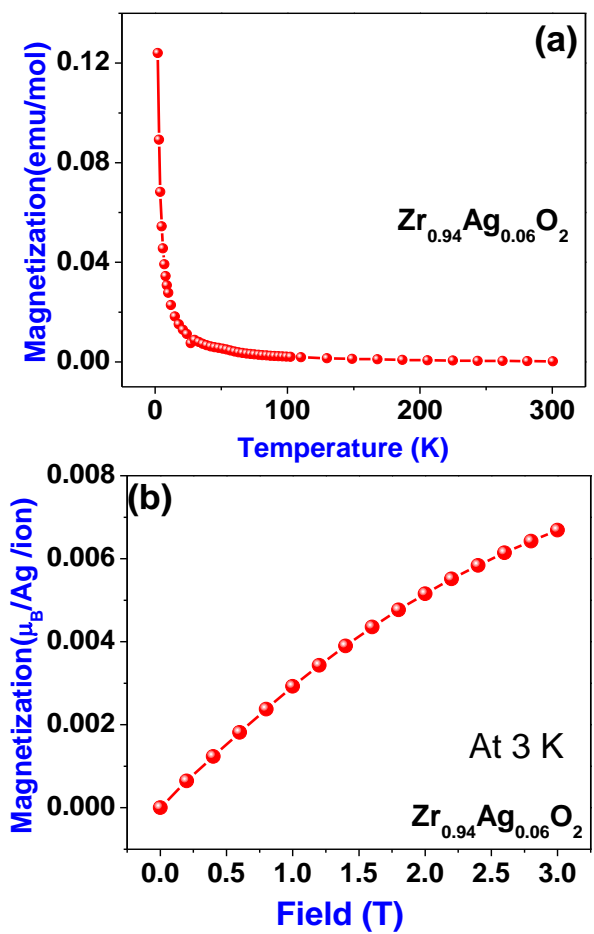

**Figure 9:** (a) Temperature variation of magnetization and (b) Magnetization versus field variation at 3K of $Zr_{0.94}Ag_{0.06}O_2$ compound, prepared at $1350^0C$.

Let us summarize the magnetism observed in these samples. The measurement of magnetic properties of these compounds indicates that pure $ZrO_2$ compound exhibit paramagnetic behavior, which is unlike few previous reports where ferromagnetism was observed in pure $ZrO_2$ [42-43]. However, the Ag-doped $ZrO_2$ compounds exhibits ferromagnetic to paramagnetic transition and FM transition temperature was found to increase with the increase of Ag concentration. Thus, the introduction of Ag in $ZrO_2$ gives rise to increase in ferromagnetic $T_C$. Moreover, the magnitude of magnetization is found to increase with Ag concentration in addition to increase in the FM $T_C$. When we consider the effect of sintering temperature on the magnetic property, the sample prepared at lower temperature is found to be ferromagnetic and this is due to the vacancies present therein.

It is an established fact in the realm of the physics and chemistry of solids that that ions substitute for one-another in structures if the charges and sizes are similar, it is usually evident from a systematic cell parameter change. In the case of a direct cationic substitution, theoretical model study [21] and first principle approach study in $ZrO_2$ [16, 26] and observation of $d^0$ magnetism in K doped $SnO_2$ [34] have demonstrated that three physical parameters are essential to explain induced $d^0$ magnetism: (i) the position of the induced impurity band (ii) the density of carrier per defect and (ii) the electrons-electrons correlations [21, 34]. In present study, $ZrO_2$ has 6-coordinate $Zr^{4+}$ and the replacement of $Zr^{4+}$ by $Ag^{1+}$ will produce three holes. To retain charge neutrality, it is possible that more defect species such as oxygen vacancies ($V_o$) are created in $ZrO_2$ structure as oxidation states of $Ag^{1+}$ is lower in comparison to $Zr^{4+}$. A vacancy induces local magnetic moments on the neighboring oxygen atoms which then interact with extended exchange couplings they interact ferromagnetically via O. Furthermore, it should be noted that we have not observed any secondary phase from the crystal structure and thus the observed ferromagnetism has intrinsic nature. It should be noted that the tetragonal or cubic symmetry of $ZrO_2$ is not absolutely necessary to obtain $d^0$ ferromagnetism.

## 4. Conclusion

To conclude, polycrystalline samples of $Zr_{1-x}Ag_xO_2$ (x=0, 0.02, 0.04, 0.06 and 0.08) were prepared by solid-state reaction route. All the prepared compounds are found to crystallize in monoclinic symmetry of $ZrO_2$ with typical lattice parameters of a=5.1468Å, b= 5.2041Å, c=5.3198 Å for pure $ZrO_2$ compound. The measurement of magnetic properties of these compounds indicates that pure $ZrO_2$ compound exhibits paramagnetic behavior. However, the Ag-doped $ZrO_2$ compounds exhibit ferromagnetic to paramagnetic transition. The Curie temperature ($\theta_C$) was found to increase from 28.7K for x=0.02 to 173.2 K for x= 0.08 doped $ZrO_2$. Thus, the introduction of Ag in $ZrO_2$ induces ferromagnetism with a large $\theta_C$. The measurement of hysteresis curves indicates that Ag doped $ZrO_2$ compounds exhibit hysteresis loops with a coercivity of around 1350 Oe. In this study, increase in Ag concentration resulted increase in the value of saturation magnetization ($M_S$). The maximum value of $M_S$ was found to be 0.009 $\mu_B$/Ag ion for x= 0.06 sample. The study of influence of sintering temperature suggests that ferromagnetism observed in the sample prepared at low temperature ($500^0C$) is possibly due to oxygen vacancies present in the sample. The sintering of sample at high temperature ($1350^0C$) diminishes the ferromagnetism and it leads to paramagnetic behaviour.

### Acknowledgements

This research did not receive any specific grant from funding agencies in the public, commercial, or not-for-profit sectors.